\documentclass[aoas,MSNbibl,nameyear,dvips]{arximspdf}

\doi{10.1214/10-AOAS409}
\referstodoi{10.1214/10-AOAS398}
\volume{5}
\issue{1}
\pubyear{2011}
\firstpage{96}
\lastpage{98}

\begin{document}
\begin{frontmatter}

\title{Discussion of:  A statistical analysis of multiple temperature proxies: Are
reconstructions of surface temperatures over the last 1000~years~reliable?}
\runtitle{Discussion}
\pdftitle{Discussion on A statistical analysis of multiple temperature proxies:
Are reconstructions of surface temperatures over the last 1000 years reliable?
 by B. B. McShane and A. J. Wyner}

\begin{aug}
\author{\fnms{Jonathan} \snm{Rougier}\corref{}\ead[label=e1]{j.c.rougier@bristol.ac.uk}}

\runauthor{J. Rougier}

\affiliation{University of Bristol}

\address{Department of Mathematics\\
University of Bristol\\
University Walk\\
Bristol BS8 1TW\\
United Kingdom\\
\printead{e1}} %adresu isvedimo komanda gale!
\end{aug}

% HISTORY:
\received{\smonth{9} \syear{2010}}

% ABSTRACT

% KEYWORDS

\end{frontmatter}

The authors are to be congratulated on the clarity of their paper,
which gives discussants and readers much to sink their teeth into.  My
comments are somewhat critical, but this should in no way devalue this
paper as an important contribution to the ongoing debate concerning
the information about historical climates that is recoverable from
proxies.  Figure~14, in particular, provides much food for thought.

In Section~3.2, comparing the proxy-based reconstruction of climate to
measures based on actual climate (in-sample mean and ARMA model) is
not very helpful for assessing the performance of the proxy---in fact,
it confirms information already presented about the nature of the
climate process and the relative variability of the proxies.  This
distracts from the more pertinent finding in Section 3.3 that the
proxy-based reconstruction seems to perform no better than various
random proxies.  Again, though, this result is not necessarily
detrimental to the proxy.  If one generates 1138 random sequences of
length 149 with roughly the right time-series properties, one should
not be surprised to find that a 1139th sequence is near the span of a
small subset, and it is a testament to the Lasso procedure that it
seems to be doing a good job at picking this subset out.  Hold-outs at
the end of the calibration period would provide a more powerful test;
for hold-outs in the middle, one can be fairly confident that if the
Lasso finds a match at both ends, then the middle will fit reasonably
well.  In Section~3.5, the finding that large numbers of
pseudo-proxies are selected can be explained in the same way.
Moreover, the Lasso procedure will have a bias against selecting
actual proxies, if they are correlated with each other.  Overall, I do
not think that Section~3 presents evidence against the proxies.

I am bemused by Section~5.  First, let us be very clear that this is
not a ``fully Bayesian'' analysis.  What we have here is a normalised
likelihood function over $\beta$ and $\sigma$ masquerading as a
posterior distribution, in order to implement a sampling procedure
over the model parameters.  This seems a perfectly reasonable
ad-hockery [although a Normal Inverse Gamma conjugate analysis
would be more conventional; see \citet{ohf04}, Chapter~11], but to call it
``fully Bayesian'' is stretching the point.  No attempt has been made
to write down a joint probability distribution over the observations
and the predictands, notably one that accounts for the possibility of
auto-correlated error in the proxy reconstruction.  Furthermore, the
reconstructions are clearly not conditional on the calibration data,
which is what the authors assert in Section 5.3.  If they were, then
there would be no reconstruction uncertainty over the calibration
period.

Then there is Figure 15, which is referred to repeatedly to show the
poor performance of the proxy-based reconstruction over the
calibration period, particularly the 1990s.  The statistical model for
this figure is initialised with temperatures from 1999 and 2000.  But
1998 was probably the warmest year of the millennium, as the authors
themselves cite in Section 1, and so the two initialisation values are
going to start the reconstruction curve too low.  What we may have
here is an artifact of a somewhat arbitrary choice of initialisation
period.  The authors must present evidence that the curve is robust to
these choices.

Finally, I have a deeper concern, not about the authors' paper in
particular, but about the general principles of reconstruction
discussed here.  There is a rich literature on statistical methods for
reconstructions; \citet{braak95} provides a review.  In this
literature, a distinction is made between the ``classical'' approach, in
which the proxies $X$ are regressed on climate quantities $Y$, and the
``inverse'' approach in which the climate quantities are regressed on
the proxies.  An advantage of the inverse approach is that it is very
tractable---it can proceed one climate quantity at a time, and it
leads to a simple plug-in approach in which the historical proxy $x_0$
is used directly to predict the historical climate value $y_0$.  The
classical approach, on the other hand, is a joint reconstruction over
several climate quantities, and requires more complicated methods to
predict $y_0$ from $x_0$, such as numerical optimisation (or a
Bayesian approach).  In its favour, however, the classical approach
respects the dominant causal direction (from climate to the proxies)
and the statistical model can reflect known features of the ecological
response function.  The broad finding regarding these two approaches
is unsurprising: the classical approach performs better in
extrapolation.  Given that historical climate reconstruction is
clearly an extrapolation from the climate in the calibration period,
and given that the proxies generally respond to multiple aspects of
climate, the use of the inverse approach, as adopted by the authors
and their forerunners, seems to me to sacrifice too much to
tractability.

\mbox{}

% imsref loaded by smiklovaite, 2011-01-19 15:13:49

\printaddresses

\end{document}